\documentclass[twocolumn] {aastex63}

\usepackage{lineno}

\begin{document}

\def\CIVdbl{{\rm C~}\kern 0.1em{\sc iv}~$\lambda\lambda 1548, 1550$}
\def\MgIIdbl{{\rm Mg~}\kern 0.1em{\sc ii}~$\lambda\lambda 2796, 2803$}
\def\NVdbl{{\rm N}\kern 0.1em{\sc v}~$\lambda\lambda 1238, 1242$}  
\def\OVIdbl{{\rm O}\kern 0.1em{\sc vi}~$\lambda\lambda 1031, 1037$}
\def\SiIVdbl{{\rm Si~}\kern 0.1em{\sc iv}~$\lambda\lambda 1394, 1403$}
\def\AlIIIdbl{{\rm Al~}\kern 0.1em{\sc iii}~$\lambda\lambda 1855, 1863$}
\def\FeIIdbl{{\rm Fe~}\kern 0.1em{\sc ii}~$\lambda\lambda 2383, 2600$}

\def\AlII{\hbox{{\rm Al~}\kern 0.1em{\sc ii}}}
\def\AlI{\hbox{{\rm Al~}\kern 0.1em{\sc i}}}
\def\AlIII{\hbox{{\rm Al~}\kern 0.1em{\sc iii}}}
\def\CaI{\hbox{{\rm Ca}\kern 0.1em{\sc i}}}
\def\CaII{\hbox{{\rm Ca}\kern 0.1em{\sc ii}}}
\def\CrII{\hbox{{\rm Cr}\kern 0.1em{\sc ii}}}
\def\C{\hbox{{\rm C~}}}
\def\CI{\hbox{{\rm C~}\kern 0.1em{\sc i}}}
\def\CII{\hbox{{\rm C~}\kern 0.1em{\sc ii}}}
\def\CIII{\hbox{{\rm C~}\kern 0.1em{\sc iii}}}
\def\CIV{\hbox{{\rm C~}\kern 0.1em{\sc iv}}}
\def\CV{\hbox{{\rm C}\kern 0.1em{\sc v}}}
\def\H{\hbox{{\rm H}}}
\def\HI{\hbox{{\rm H~}\kern 0.1em{\sc i}}}
\def\HII{\hbox{{\rm H~}\kern 0.1em{\sc ii}}}
\def\Lya{\hbox{{\rm Ly}\kern 0.1em$\alpha$}}
\def\Lyb{\hbox{{\rm Ly}\kern 0.1em$\beta$}}
\def\Lyg{\hbox{{\rm Ly}\kern 0.1em$\gamma$}}
\def\Lyfive{\hbox{{\rm Ly}\kern 0.1em$5$}}
\def\Lysix{\hbox{{\rm Ly}\kern 0.1em$6$}}
\def\Lyseven{\hbox{{\rm Ly}\kern 0.1em$7$}}
\def\Lyeight{\hbox{{\rm Ly}\kern 0.1em$8$}}
\def\Lynine{\hbox{{\rm Ly}\kern 0.1em$9$}}
\def\Lyten{\hbox{{\rm Ly}\kern 0.1em$10$}}
\def\HeI{\hbox{{\rm He}\kern 0.1em{\sc i}}}
\def\HeII{\hbox{{\rm He}\kern 0.1em{\sc ii}}}
\def\FeI{\hbox{{\rm Fe~}\kern 0.1em{\sc i}}}
\def\FeII{\hbox{{\rm Fe~}\kern 0.1em{\sc ii}}}
\def\FeIII{\hbox{{\rm Fe~}\kern 0.1em{\sc iii}}}
\def\MnII{\hbox{{\rm Mn}\kern 0.1em{\sc ii}}}
\def\MgI{\hbox{{\rm Mg~}\kern 0.1em{\sc i}}}
\def\MgII{\hbox{{\rm Mg~}\kern 0.1em{\sc ii}}}
\def\MgIII{\hbox{{\rm Mg~}\kern 0.1em{\sc iii}}}
\def\MgIV{\hbox{{\rm Mg~}\kern 0.1em{\sc iv}}}
\def\NaI{\hbox{{\rm Na}\kern 0.1em{\sc i}}}
\def\NV{\hbox{{\rm N}\kern 0.1em{\sc v}}}
\def\NII{\hbox{{\rm N}\kern 0.1em{\sc ii}}}
\def\NIII{\hbox{{\rm N}\kern 0.1em{\sc iii}}}
\def\O{\hbox{{\rm O}}}
\def\Mg{\hbox{{\rm Mg}}}
\def\Fe{\hbox{{\rm Fe}}}
\def\OVI{\hbox{{\rm O}\kern 0.1em{\sc vi}}}
\def\OIV{\hbox{{\rm O}\kern 0.1em{\sc iv}}}
\def\OI{\hbox{{\rm O}\kern 0.1em{\sc i}}}
\def\OII{\hbox{{\rm O}\kern 0.1em{\sc ii}}}
\def\OIII{\hbox{{\rm O}\kern 0.1em{\sc iii}}}
\def\PV{\hbox{{\rm P}\kern 0.1em{\sc v}}}
\def\SiII{\hbox{{\rm Si~}\kern 0.1em{\sc ii}}}
\def\SiIII{\hbox{{\rm Si~}\kern 0.1em{\sc iii}}}
\def\SiIV{\hbox{{\rm Si~}\kern 0.1em{\sc iv}}}
\def\SII{\hbox{{\rm S}\kern 0.1em{\sc ii}}}
\def\SIII{\hbox{{\rm S}\kern 0.1em{\sc iii}}}
\def\SIV{\hbox{{\rm S}\kern 0.1em{\sc iv}}}
\def\SVI{\hbox{{\rm S}\kern 0.1em{\sc vi}}}
\def\TiII{\hbox{{\rm Ti}\kern 0.1em{\sc ii}}}
\def\ZnII{\hbox{{\rm Zn}\kern 0.1em{\sc ii}}}
\def\kms{\hbox{km~s$^{-1}$}}      
\def\cmsq{\hbox{cm$^{-2}$}}
\def\cc{\hbox{cm$^{-3}$}}
\newcommand{\etal}{et~al.\ }
\newcommand{\minfit}{\sc minfit}
\newcommand{\DR}{\hbox{\sc dr}}
\newcommand{\s}{$\sigma$}
\def\lb{$\lambda$}
\def\ang{$\text \AA$}

\def\lsim{\mathrel{\rlap{\lower4pt\hbox{\hskip1pt$\sim$}}
    \raise1pt\hbox{$<$}}}                
\def\gsim{\mathrel{\rlap{\lower4pt\hbox{\hskip1pt$\sim$}}
    \raise1pt\hbox{$>$}}}                
\def\Msun{\hbox{M$_{\odot}$}}

\title{Connection between Emission and Absorption Outflows through the Study of Quasars with Extremely-High Velocity Outflows}

\author{Paola Rodr\'iguez Hidalgo}
\affiliation{Physical Sciences Division -- School of STEM \\
University of Washington Bothell \\
Bothell WA, 98011, USA}

\author{Amy L. Rankine}
\affiliation{Institute for Astronomy, University of Edinburgh \\
Royal Observatory, Blackford Hill \\
Edinburgh, EH9 3HJ, UK}

\begin{abstract}
A recently-discovered class of outflows,  extremely high-velocity outflows (EHVOs), may be key to understanding feedback processes as it is likely the most powerful in terms of mass-energy.
These EHVOs have been observed at redshifts 1.052~$< z_{\rm em} <$~7.641, but the potential connection with outflows in emission had not been studied.
We find that EHVOs, albeit their small numbers at the moment, appear to show distinct {\CIV} and {\HeII} properties. In particular, EHVOs are more predominant in quasars with large blueshifts of the {\CIV} emission line, suggesting a connection between emission and absorption
outflowing signatures for these extreme outflows. We also find incipient trends with the maximum velocity of the outflows, which is similar to what has been previously found in BALQSOs, but now extending previous studies to speeds up to $\sim$0.2$c$.
We find the bolometric luminosities, Eddington ratios, and black hole masses of our sample are overall very similar from the general quasar population upon considering their {\CIV} emission properties.
This is close to the case for {\HeII} EW as we observe a tentative upper limit to the {\HeII} strength for a quasar to host an EHVO.
This study shows that extreme outflows such as EHVOs appear in quasars that are clearly a distinct class from the overall BALQSO population, and solidify the relation between outflows observed in emission and in absorption. 

\end{abstract}


{

\section{Introduction} \label{sec:intro}

Quasars, the most luminous of the active galactic nuclei (AGNs), are found at the center of the most massive galaxies (e.g., \citealt{Lynden-Bell69}; \citealt{Bahcall97}).  
Quasars' large luminosities allow us to study them at large redshifts, providing information about galactic evolution within our universe. 

Outflows are fundamental constituents of AGNs and they provide first-hand information about the physical and chemical properties of the AGN environment. 
They are studied in emission through the analysis of
blueshift of emission lines (e.g., \citealt{Komossa15}; \citealt{Marziani17}) and through absorption-line signatures (e.g., broad, blue-shifted resonance lines in the UV and X-ray bands) as the gas intercepts some of the light from the central continuum source and broad emission-line region (e.g. \citealt{Crenshaw99}; \citealt{Reichard03}; \citealt{Hamann04}; \citealt{Trump06}; \citealt{Dunn08}; \citealt{Ganguly08}; \citealt{Nestor08} and references therein).  
Outflows have been invoked as a potentially regulating mechanism that would provide the necessary energy and momentum ``feedback'' \citep[e.g.,][]{Silk98, DiMatteo05, Springel05, Hopkins06} required to explain the correlation between the black hole masses ($M_{\rm BH}$) and the masses of the stellar spheroids ($M_{\rm bulge}$) of their host galaxies  \citep[e.g.,][]{Gebhardt00, Merritt01, Tremaine02}.

Outflows at large speeds are likely to be the most effective way of transporting energy from the AGN to galactic scales, therefore having more influence on the formation and destruction of galactic structures (\citealt{Silk98}; \citealt{Scannapieco04}; \citealt{Hopkins06}). Outflows with speeds $v\sim$0.2$c$ may carry 1--2.5 orders-of-magnitude larger kinetic power than gas outflowing at what is defined as ``high'' velocities ($v \sim$5,000--10,000 {\kms}), assuming similar distances from the inner source and similar physical properties of the gas, because kinetic power is proportional to $v^3$. 
Such outflows have been detected both in X-ray and UV/optical spectra. In the X-rays, the so-called Ultra Fast Outflows (UFOs) have been observed as Fe K-shell absorption in the X-ray spectra of nearby AGNs (predominantly Seyferts) at speeds similar and even larger than those in EHVOs (0.03$c$--0.4$c$; e.g., \citealt{Chartas02}; \citealt{Reeves03}; \citealt{Tombesi10}). However, UFOs have been rarely detected in high-$z$ quasars: there are just over 10 cases with $z_{\rm em} > $1.5 to date; a systematic study of UFOs at large redshifts is prohibitive at the moment.

The number of known UV/optical broad (FWHM $\gsim$ 1,000 {\kms}) EHVOs
in quasar spectra has exploded in the last two years from only a handful of cases in previous years (\citealt{Januzzi96}; \citealt{Hamann97a}; \citealt{RodriguezHidalgo11}; \citealt{Rogerson16}), to more than 150 new cases. We  have discovered 138 of these new cases at 1.9~$< z \lsim$~4.5:
40 cases (\citealt{RodriguezHidalgo20}) in the ninth release of the Sloan Digital Sky Survey (SDSS) quasar catalog (DR9Q; \citealp{Paris12}), and 98 new cases (Rodr\'iguez Hidalgo et al.~in prep.) in the DR16Q (\citealt{Lyke20}).  
As it is common practice in surveys of BALQSOs (e.g., \citealt{Gibson09b}), we used the presence of {\CIV} $\lambda\lambda$1548.1950,1550.7700 absorption that appears blue-shifted relative to the {\CIV} emission to identify the EHVO. Among different ionic transitions, {\CIV}  
is (1) commonly present in quasar outflows, and (2) easily observed due to the fact that it is redshifted into the optical range for quasars in the epoch of peak quasar activity (for luminous Type 1 quasars, the comoving space density peaked at redshifts 2$< z_{\rm em} <$3; \citealt{Schmidt95}; \citealt{Ross13} and references therein). 
Prior surveys of {\CIV} BALQSOs had set an arbitrary upper velocity limit of 0.1$c$ to avoid complications due to misidentification with {\SiIV} and other ionic transitions bluewards of the {\SiIV} emission line, but we have developed a method that flags all absorption in the spectral region of interest, and allows the user to visually inspect the spectrum easily to help identify the right {\CIV} transitions, rejecting or including the flagged candidate as a confirmed EHVO case. Narrow (FWHM $\lsim$ 500 {\kms}) EHVOs have also been studied in statistical studies (e.g., \citealt{Nestor08}; \citealt{Chen21}), where properties like line locking or correlations with the presence of other outflows suggest an intrinsic nature. 
 
The connection between emission and absorption outflows at these extreme speeds is completely unknown, but there are previous studies in BALs at lower speeds. 
Using data in DR7, \citet{Richards11} found a relationship between {\CIV} emission line properties and whether there was absorption present in quasar spectrum by studying a sample of $\sim$30,000 quasar spectra. They found a dearth of BALQSOs, which were defined as having broad absorption at speeds up to 20,000 {\kms}, in quasars that show {\CIV} emission lines with small blueshifts and large equivalent widths.
Most recently, \citet{Rankine20} has extended this work to 144,000 quasars in SDSS DR14Q and studied the effects of BAL properties. We found that the Balnicity Index \citep[BI;][]{Weymann91} and other BAL trough parameters, including the maximum and minimum trough velocities, all increase as the {\CIV} blueshift increases. Also of note is the finding that BAL and non-BAL quasars occupy the same {\CIV} emission space, apart from the region of highest EW and low blueshifts where there are few BALQSOs. The fraction of quasars that have BALs, however, changes across the space with the highest BAL fraction occurring at the highest blueshifts. Quasars with similar {\CIV} emission profiles were found to have similar optical luminosities, Eddington ratios, and, most notably, {\HeII} $\lambda$ 1640 EW which is an indicator of the strength of the ionising SED \citep{Leighly04}. Other studies have also found the EW of {\HeII} to anti-correlate with {\CIV} emission blueshift \citep{Baskin13, Baskin15} 
and \citet{Rivera22} has shown that {\HeII} EW is linearly anti-correlated with {\CIV} \textit{distance}, a metric describing the distance along a best-fit curve through the {\CIV} space: values of {\CIV} distance are zero at high-EW and low-blueshift (top left) and they increase towards unity along the best-fit curve as we move towards low-EW and high-blueshift (bottom right). The best-fit curve was constructed by \citet{CIVdist} with piecewise polynomial fitting to the \citet{Rankine20} sample. Each quasar can then be projected onto the best-fit line and a {\CIV} distance value is assigned.
This metric was first introduced by \citet{Rivera20} to better explore the variability of the SDSS Reverberation Mapping quasars \citep{Shen15} and by \citet{Richards21} to investigate radio trends in {\CIV} space. \citet{Rivera22} suggest that {\CIV} distance can be used as a proxy for the Eddington ratio and so the {\HeII} EW (thus the SED hardness) is anti-correlated with the Eddington ratio. 

However, both \citet{Richards11} and \citet{Rankine20} were limited to absorption outflowing up to speeds of 20,000--25,000 {\kms}, as most BALQSO studies are. EHVOs present a test to any trend found in BALQSOs as we can triple the range of speeds we can study up to 60,000 {\kms}.

In this paper, we study whether there is a connection between outflows in {\CIV} emission and extreme outflows found in absorption as EHVOs. 
In section \S\ref{sec:data} we discuss the origin of the known properties we cross-correlate, and in \S\ref{sec:results} we study potential connections between {\CIV} outflows in absorption and emission by studying the relation between properties of emission lines and our measured properties of EHVOs, as well as contrasting EHVOs in relation to BALQSOs. We discuss their implications in\S\ref{sec:discussion}.

\section{Data} \label{sec:data}

The data on the DR9Q 40 EHVO quasars  was published in \citet{RodriguezHidalgo20} (hereafter, RH+2020), where we selected quasars with $z_{\rm em} \geq$~1.9 and signal-to-noise ratio ($S/N$) larger than 10, and removed cases without sufficient wavelength coverage in the region of interest, resulting in a parent sample of 6743 quasars. Spectra were then normalized and we systematically searched for EHVO absorption. 

The measurements of {\CIV} emission line blueshifts and EWs, as well as {\HeII} EW, used in this paper were measured as part of the analyses of \citet{Rankine20} (hereafter, Rankine+2020). We produced spectrum reconstructions covering restframe 1260--3000\,\AA\ generated by a mean field independent component analysis of a selection of 144,000 quasars in the DR14Q \citep{Paris18} with 1.5~$\lsim z_{\rm em} \leq$~3.5, 45.3~$< \log_{10}(L_{\rm bol}) <$~48.2 erg s$^{-1}$, and spectra with an average $S/N \geq 5.0$ per SDSS spectrum pixel.
The reconstructions allow for measurements of the intrinsic {\CIV} emission to be made even in the presence of extensive absorption in BAL quasars and also reduce the uncertainty in these measurements in low-S/N spectra by using the information across the entire spectrum to inform the reconstruction. Use of the reconstructions and the non-parametric nature of this procedure remove the requirement to fit Gaussian profiles to the spectra. We refer the reader to section 6.1 in Rankine+2020 for a detailed description but, in summary, calculation of the {\CIV} emission line parameters followed a non-parametric approach: first, a power-law continuum was substracted from the reconstruction, then the emission-line flux was integrated to determine EW, and the blueshift was calculated from the wavelength which bisects the total cumulative line flux, relative to the systemic velocity of the quasar.

\begin{figure}
\centering
\includegraphics[width=1.0\linewidth]{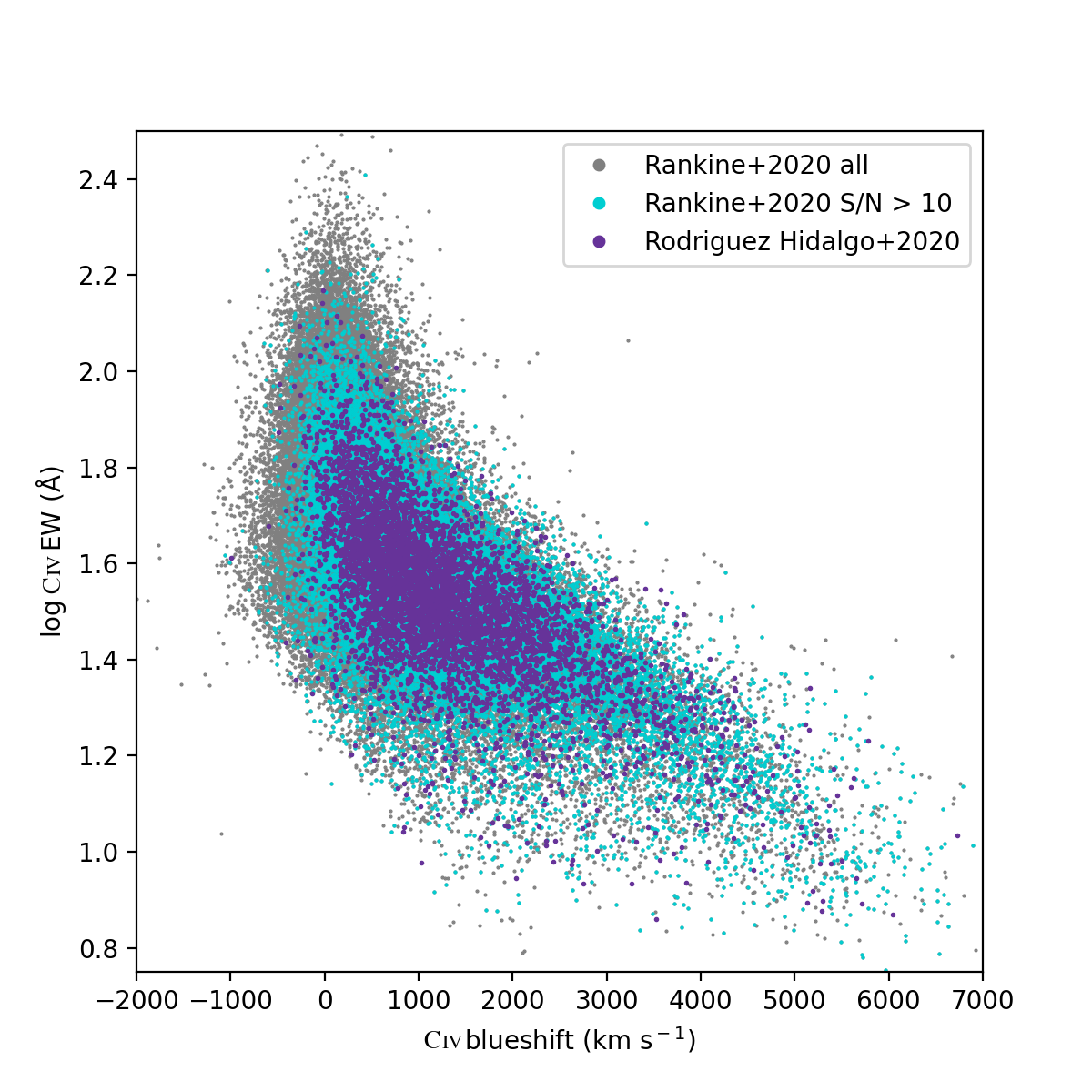}
\caption{Comparison of \citet{Rankine20} and \citet{RodriguezHidalgo20} samples in the studied parameter space. 
The lack of overlap in the top-left quadrant is due to the different signal-to-noise ratios set in both samples, but it is reduced once we establish the same cutoff. 
}
\label{CIVblue}
\end{figure}
 
We cross-correlated these two samples using the observation information, namely plate-mjd-fiber, to study where EHVOs lie in the {\CIV} blueshift - EW parameter space. The redshift cutoff in Rankine+2020\footnote{Systemic redshifts were calculated using an independent component analysis (ICA) of the 1600--3000\,\AA\ restframe wavelength region \citep{Allen13} with redshift as a free parameter. Please see Rankine+2020 for an in-depth description, but in summary the 1600--3000\,\AA\ region is chosen deliberately to involve low- as opposed to high-ionization lines and
thereby exclude the often blueshifted {\CIV} emission line which can bias the redshifts by $>1000$ {\kms} for the highest {\CIV} blueshift sources. 
Indeed, the ICA-based redshifts are much more consistent with the redshifts of \citet{Wu_catalog_2022} for SDSS DR16 than the original SDSS redshifts.
We use these ICA redshifts to recalculate EHVO speeds so they are slightly different than in RH+2020; however, we note that our results do not change qualitatively if we use the EHVO speeds from RH+2020.} only excludes 3\% of our parent sample (215/6743) but 20\% (8/40) of our quasars with EHVOs; this was expected as EHVOs seem to be more predominant at higher redshifts (RH+2020; \citealt{Bischetti22}; \citealt{Wang18}; \citealt{Wang21}). 
We contrasted the parent samples to make sure that differences between them would not affect our study. We observed that the largest difference was due to the cutoffs in signal-to-noise ratio: average $\rm S/N \geq 5.0$ in Rankine+2020 versus $\rm S/N >$~10 in RH+2020. 
Once we apply the same cutoff in the Rankine+2020's sample both samples mostly overlap (see Fig.~\ref{CIVblue}). 
Together with the restriction of good spectrum reconstructions with reduced-$\chi^2<2$ and excluding the lower-left corner of {\CIV} space which contains mostly intrinsically pathological spectra and FeLoBALs for which the ICA reconstruction scheme is not designed (details in section 4.3 of Rankine+2020), 
the parent sample in Rankine+2020 is reduced to 41,535 quasars (aqua in Fig.~\ref{CIVblue}), the parent sample in RH+2020 to 5,730 quasars (darker blue in Fig.~\ref{CIVblue}), and the number of EHVO quasars to 31. 
The resulting luminosity range in Rankine+2020 is similar to that for the parent sample in RH+2020. 
Measurements of bolometric luminosity ($L_{\rm bol}$), black hole mass ($M_{\rm BH}$) and Eddington rate ($L_{\rm bol}/L_{\rm Edd}$) were also taken from Rankine+2020. Only values of $M_{\rm BH}$, and therefore also $L_{\rm bol}/L_{\rm Edd}$, calculated for cases with {\CIV} blueshift $>$ 500 {\kms} are included because the correction to take into account the excess, non-virial, blue emission for quasars with large {\CIV} blueshifts (\citealt{Coatman17}) is not well defined for cases with negative or modest positive blueshifts. Please see section 6.3 in Rankine+2020 for more information.

We also use the {\HeII} properties from Rankine+2020 to compare the SEDs of the EHVOs to the parent sample 
in {\CIV} space and with {\CIV} distance by employing the {\tt CIVdistance} code by \citet{CIVdist} to project our sample onto the curve and measure the {\CIV} distance.

In our analysis, we also contrast quasars with EHVOs to BALQSOs. The latter were defined slightly differently in both samples. For simplicity, we use the definition in Rankine+2020: Balnicity index in {\CIV} larger than zero. BALQSOs in this sample show {\CIV} absorption with speeds up to 25,000 {\kms}.

\section{Results} \label{sec:results}

In Figure \ref{CIVbluehist}, we show that EHVOs (purple; from RH+2020) appear clearly in quasar spectra with larger values of blueshift and smaller values of equivalent width (EW) of the {\CIV} emission line than the other two populations, namely non-BALQSOs (aqua) and BALQSOs (light blue), both from Rankine+2020. 

\begin{figure}[ht!]
\centering
\includegraphics[width=1.0\linewidth]{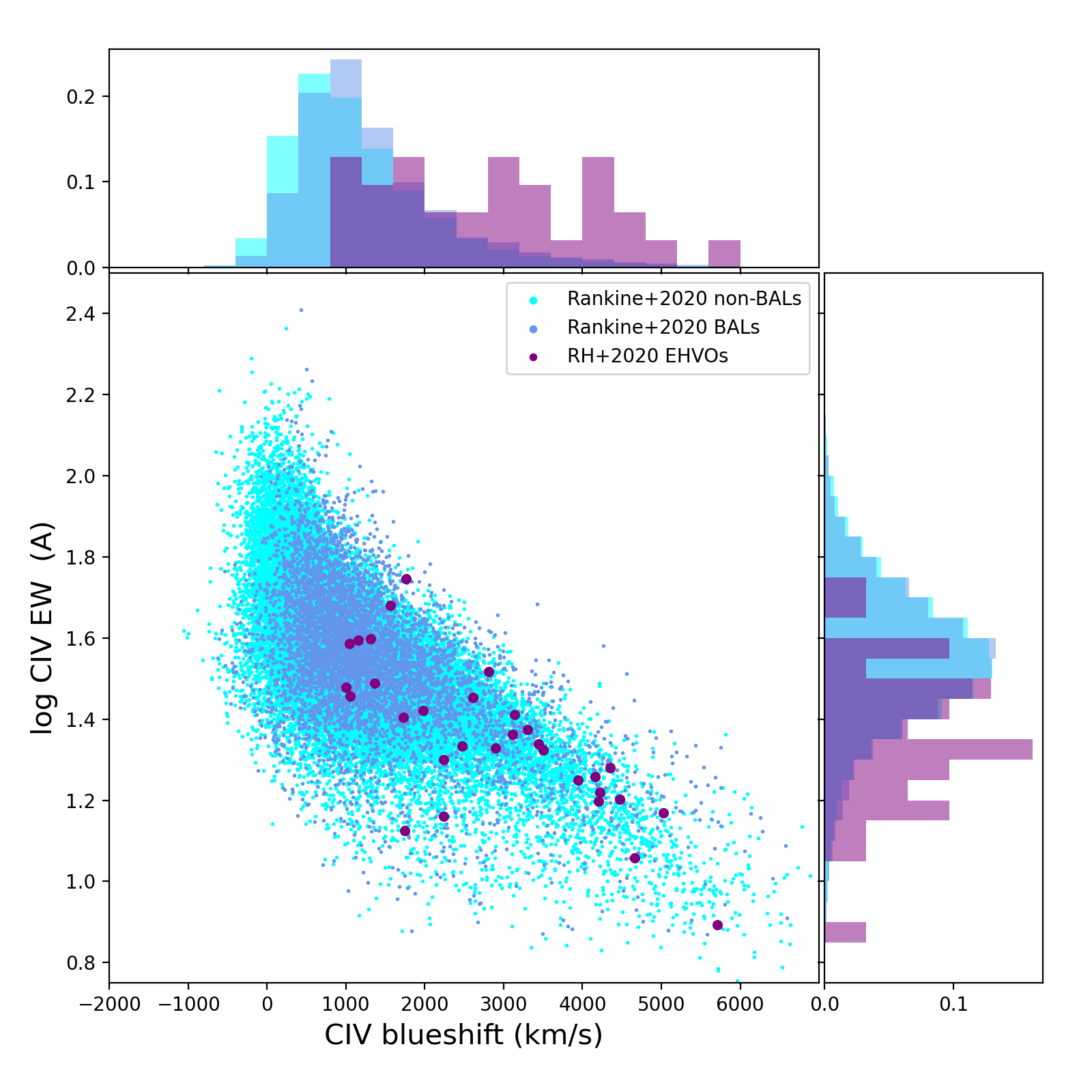}
\caption{Scatter plot shows non-BALs (aqua), BALs (light blue), both from Rankine+2020, and EHVOs from RH+2020 (purple) in the {\CIV} emission line blueshift and EW parameter space. A progression towards larger blueshifts and smaller EW is shown from the samples of non-BALs and BALs towards the EHVOs. 
}
\label{CIVbluehist}
\end{figure}

\begin{figure*}[ht!]
    \centering
    \includegraphics[width=1.\linewidth]{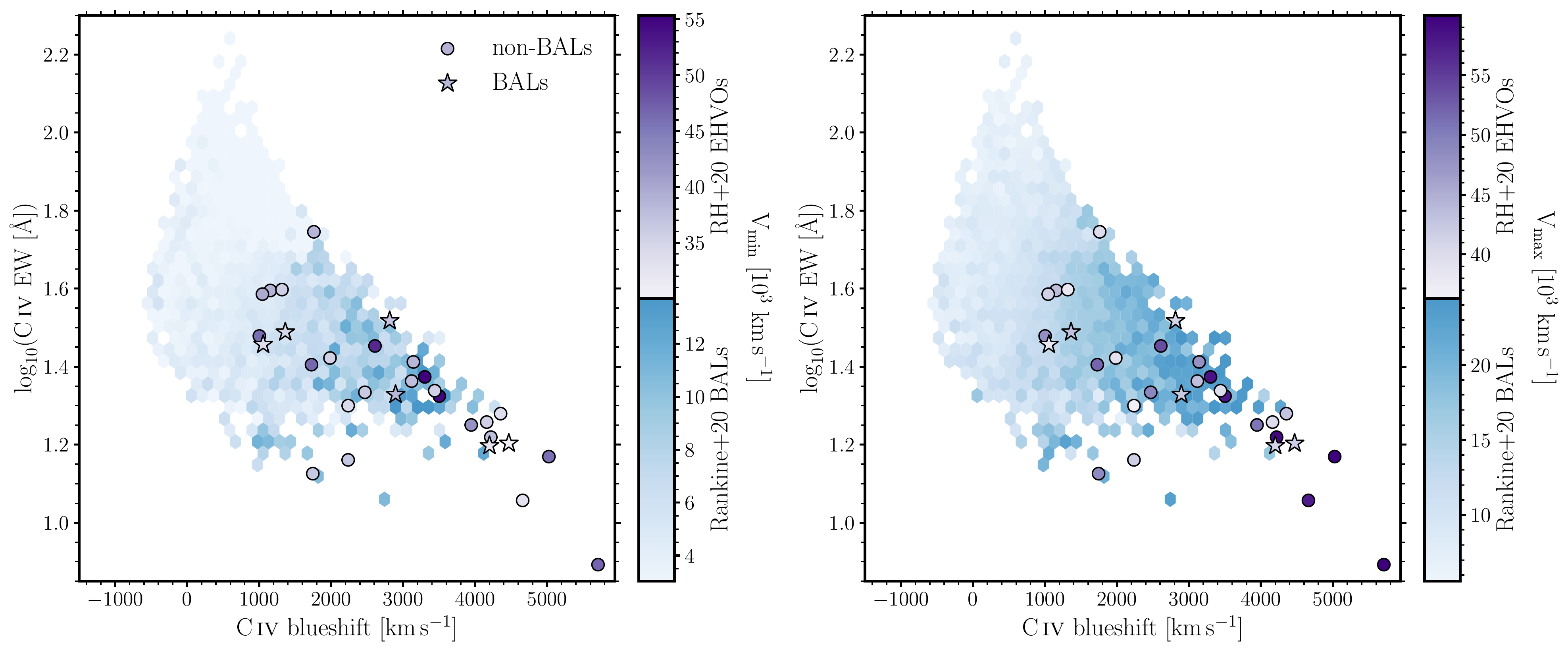}
    \caption{{\CIV} emission space colour-coded by minimum (left) and maximum velocities (right) of the Rankine+2020 BALs (hexagons; blue colour map) and EHVOs (circles; purple colour map). Note that, due to the definition of an EHVO, the minimum and maximum EHVOs are significantly greater than their BAL counterparts to warrant separate colour maps. The EHVO sample is split into BALs (stars) and non-BALs (circles). The hexagonal binning is a reproduction of Rankine+2020's figure 16: median velocities of the objects within each hexagon are calculated and only hexagons with five or more objects are plotted. The BAL minimum and maximum velocities both clearly increase with increasing {\CIV} blueshift. The maximum EHVO velocity also appears to increase; however, the minimum EHVO velocity shows no such trend. Notice also that EHVOs seem to occupy the same parameter space as BALs with speeds in the upper half ($V_{\rm max} \gsim$~15,000 {\kms}).
    }
    \label{vminvmax}
\end{figure*}

While our number of cases in each sample differ by orders in magnitude (35,694 non-BALs and 5,841 BAL quasars in Rankine+2020, and 31 EHVO quasars in RH+2020), the distributions of {\CIV} emission line properties differ largely. It is implausible that 31 cases selected randomly from either of the other two distributions would result in that same distribution. We performed Kolmogorov–Smirnov (K-S) tests 
between the distribution of ${\CIV}$ blueshift values in the parent samples, the non-BALQSOs, the BALQSOs versus the EHVO sample, obtaining $p$-values smaller than 1e-08 for all the tests, with KS statistic values larger than 0.5, therefore rejecting the hypothesis that the values for EHVO come from any of the other distributions. We also tried selecting 10,000 times a random sample of 31 quasars from each of 
the comparison distributions, and counted how many times the median of this selection would be equal or larger than the median of the EHVO {\CIV} blueshift values; we obtained 0 for all cases. 

In Figure~\ref{vminvmax} we examine the relationship between minimum and maximum  velocities in {\CIV} emission space against a backdrop of the BI-defined BALQSOs trough parameters from Rankine+2020. 
Rankine+2020 found the minimum and maximum BAL velocities to correlate with {\CIV} blueshift (see also Rankine+2020's Figure 16).
Particularly noticeable from both panels is that the EHVOs, which by definition have large $V_{\rm min}$ and $V_{\rm max}$, occupy the same region of {\CIV} space where the BALQSOs also have large $V_{\rm min}$ and $V_{\rm max}$, both EHVOs and BALQSOs showing {\CIV} blueshifts $\gsim$~1,000 {\kms}. Looking now at the specific EHVO velocities, the minimum EHVO velocity does not appear to show any trend with {\CIV} blueshift. Note that the BALQSOs with EHVOs have absorption present at lower velocities that is not included in the calculation of $V_{\rm min}$ and subsequent colour-coding of the stars in the left panel of the Figure. The seven BAL quasars in the EHVO sample have BAL minimum velocities greater than 11,000 {\kms} which is consistent with their location in {\CIV} space. Due to the small sample size, there is only a tentative trend of increasing maximum EHVO velocity with increasing {\CIV} blueshift. If confirmed, this trend would be in agreement with the BALQSOs, only at much greater velocities as a result of the BAL and EHVO definitions. Extending this work to the DR16 sample will hopefully provide such a clarification. 

Figure \ref{LbolMBHRedd} shows the studied {\CIV} emission space for the parent and EHVO samples as a function of values of $L_{\rm bol}$ (left), $M_{\rm BH}$ (center), and $L_{\rm bol}/L_{\rm Edd}$ (right). Except for some outliers, it does not seem that the overall values of these three physical properties are significantly distinct for EHVOs relative to the parent sample in the {\CIV} emission space; in other words, while EHVOs show larger values overall of $L_{\rm bol}$ and $L_{\rm bol}/L_{\rm Edd}$ (see Discussion), they do not have different values of $L_{\rm bol}$, $M_{\rm BH}$, and $L_{\rm bol}/L_{\rm Edd}$ than the values expected for a quasar with those {\CIV} emission line properties.

\begin{figure*}[ht!]
\centering
\includegraphics[width=0.325\linewidth]{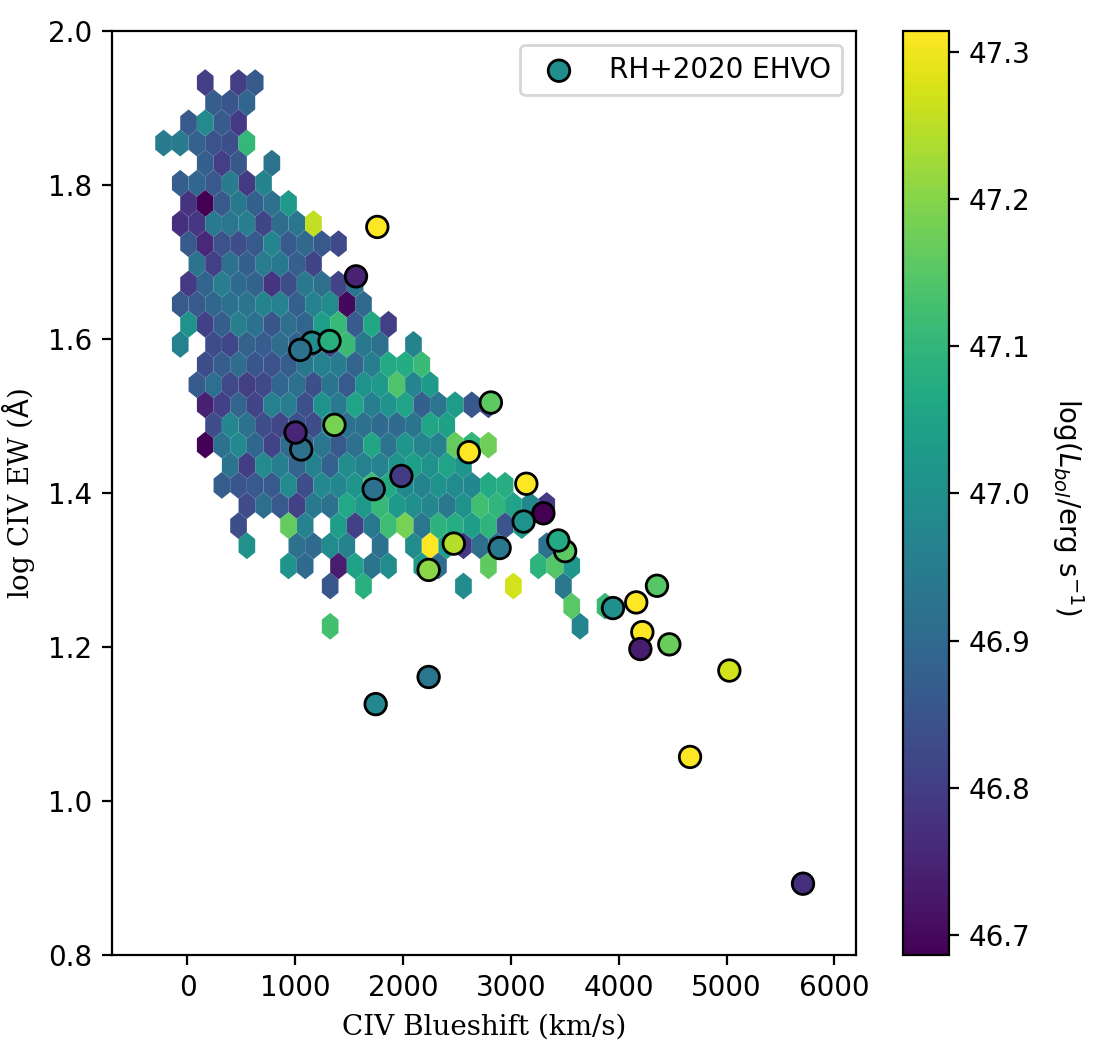}
\includegraphics[width=0.325\linewidth]{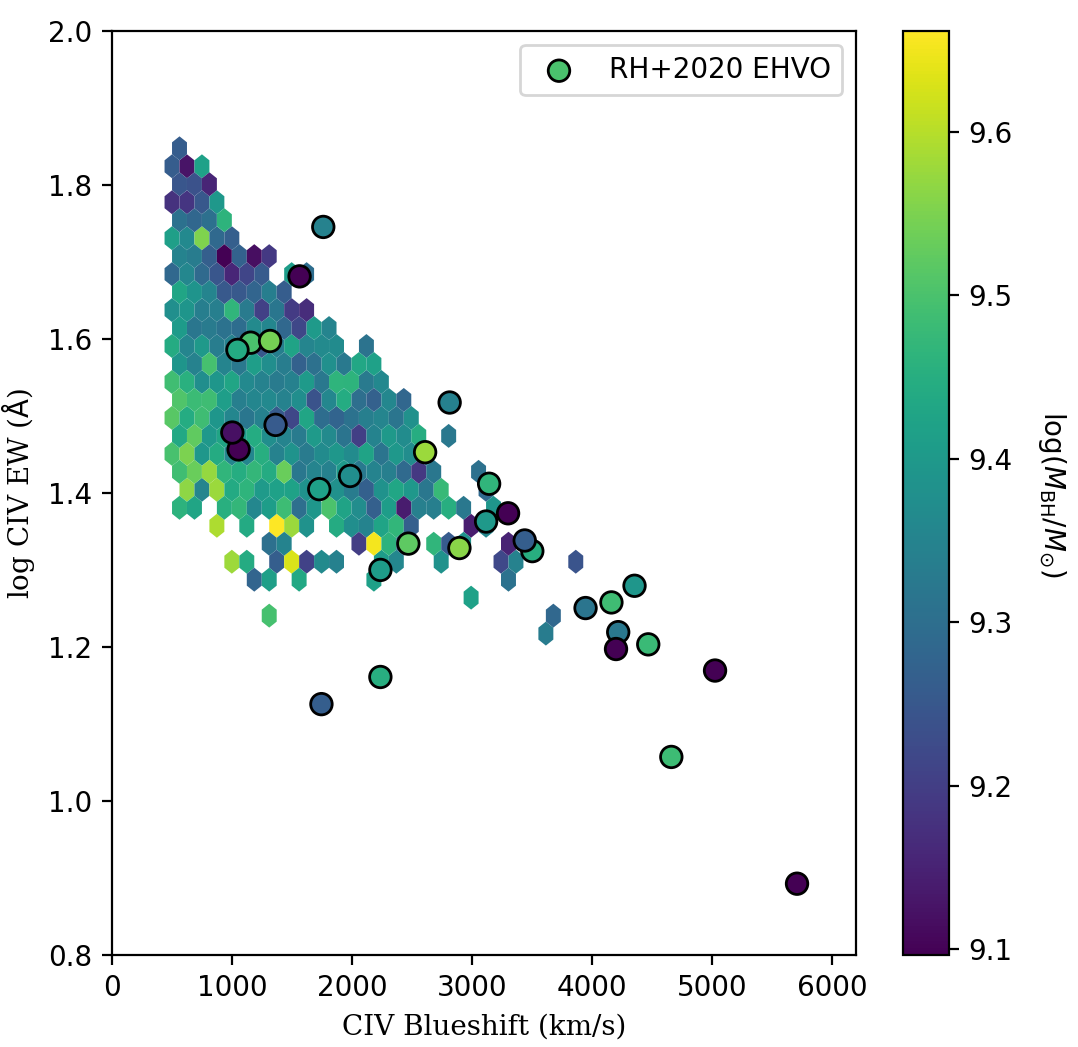}
\includegraphics[width=0.325\linewidth]{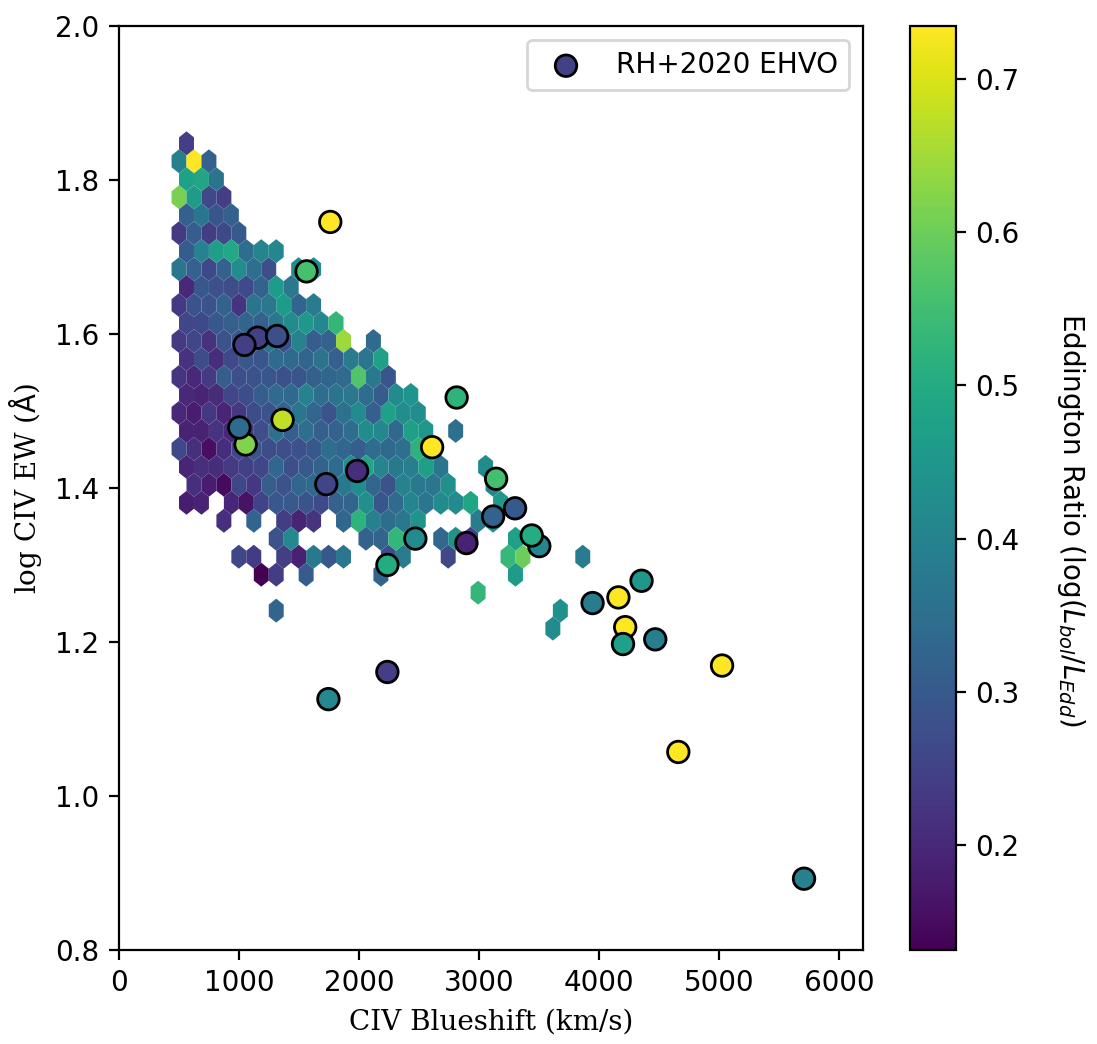}
\caption{{\CIV} emission space colour-coded by minimum to maximum values of bolometric luminosity (left), black hole mass (center) and Eddington ratio, (right), for both the DR9Q parent sample (hexagons) and EHVOs (circles). The hexagonal binning is modified due to the smaller numbers in the parent sample relative to Rankine+2020: median values of the objects within each hexagon are calculated and only hexagons with four or more objects are plotted. Notice that values of black holes mass, and therefore Eddington ratio, are only included for cases with {\CIV} blueshift $>$ 500 {\kms} because the correction used here and described in \citet{Coatman17} is not well defined for cases with negative or modest positive blueshifts (please see \S\ref{sec:data} for information about measurements).
Except for some outliers, most EHVOs are not distinguishable from the parent sample based on these three properties. 
}
\label{LbolMBHRedd}
\end{figure*}

Figure \ref{HeII_CIVdist} shows our analysis of {\HeII} EW in EHVOs. Figure \ref{HeII_CIVdist} (top) contains the {\CIV} space colour-coded by the {\HeII} EW for the EHVO and parent samples. In a similar manner to the bolometric luminosity, Eddington ratio and black hole mass, the {\HeII} EWs of the EHVOs are consistent with the parent sample and determined by their location in {\CIV} space.
There is a subtle suggestion that the EHVOs with the largest {\CIV} EW have perhaps slightly higher {\HeII} EW than would be expected based on their location in {\CIV} space. To assess this further we reduce the dimensionality by plotting the {\HeII} EW as a function of the {\CIV} distance in the bottom panel of Figure \ref{HeII_CIVdist}. {\CIV} distance has been shown to anti-correlate almost linearly with {\HeII} EW \citep{Rivera22}, so we can use this metric to examine trends in {\CIV} space more simply. 
The EHVO sample is clearly shifted to large {\CIV} distances and lower {\HeII} EW; however, from this Figure one can also see a cluster of EHVOs around a {\CIV} distance of 0.7 which have significant {\HeII} emission. The two EHVOs with the largest {\HeII} EW ($\sim$3 \AA) in our sample have {\HeII} emission that is as strong as appears to be allowed for their {\CIV} distance values based on the parent population. 
Studies with larger samples of EHVOs will facilitate further investigation of this possible phenomenon.

\begin{figure}[ht!]
\centering
\includegraphics[width=1.0\linewidth]{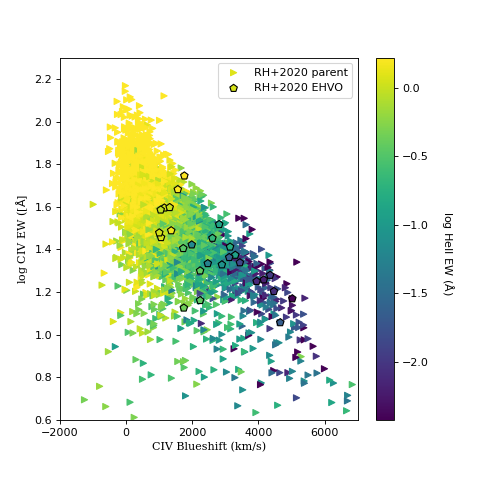}
\includegraphics[width=1.0\linewidth]{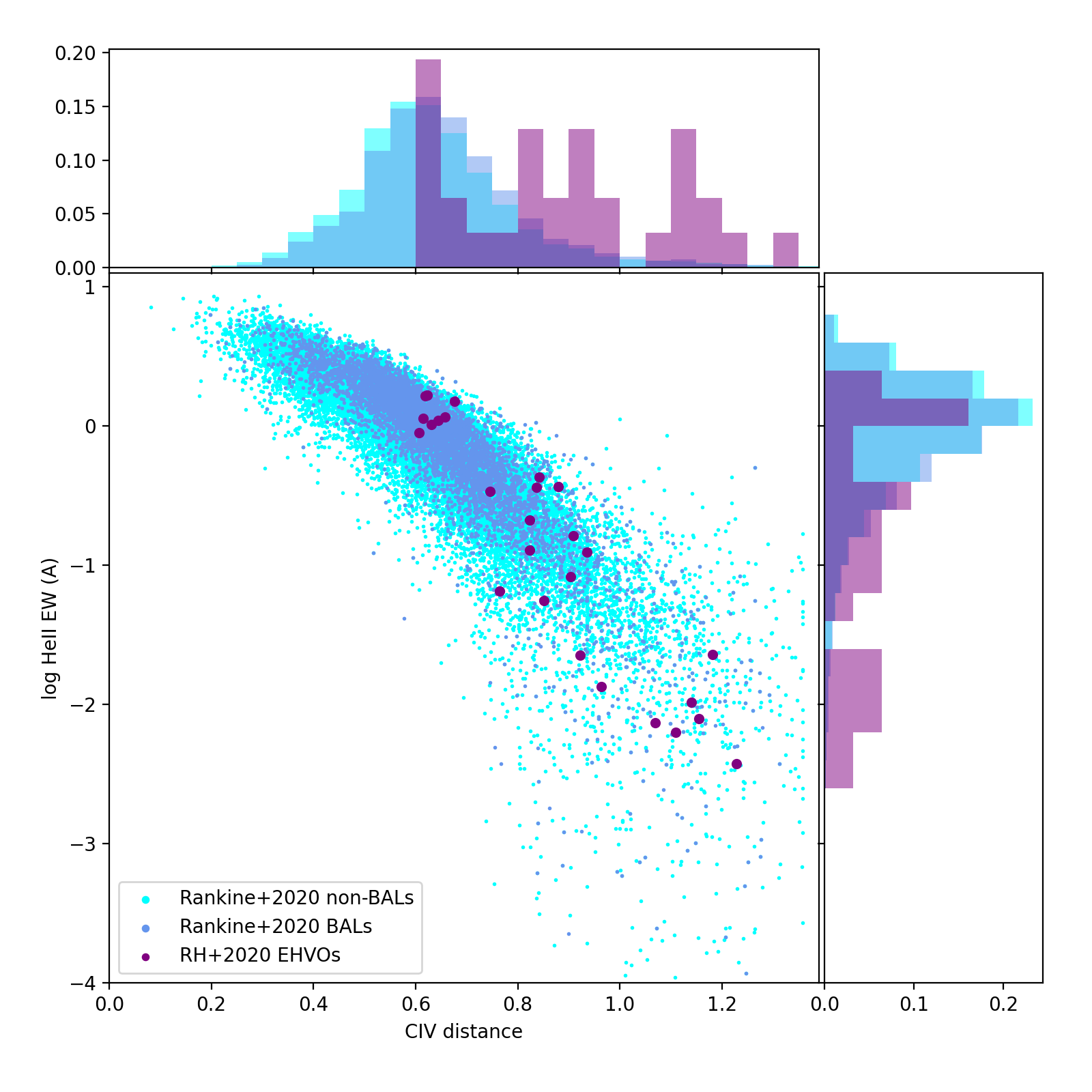}
\caption{Top: {\CIV} emission space colour-coded by {\HeII} EW. In agreement with the parent sample, the EHVO {\HeII} EW decreases as {\CIV} blueshift increases and EW decreases. The EHVOs with the largest {\HeII} EWs have perhaps too strong {\HeII} for their location in {\CIV} space. Bottom: {\HeII} EW against {\CIV} distance. The EHVO quasars are shifted to the bottom right compared to the BAL and non-BALQSOs while the EHVOs with the highest {\HeII} EWs appear to be at the limit of allowed {\HeII} values.}
\label{HeII_CIVdist}
\end{figure}

\section{Discussion} \label{sec:discussion}

The literature is only sparsely populated with investigations connecting the outflows observed in absorption with those in emission at intermediate redshifts. Primarily because parameterising the {\CIV} emission outflow signature is challenging in quasars with extensive {\CIV} absorption. For example, \citet{Richards11} used the blueshifted {\CIII}] complex as a proxy for {\CIV} blueshift; however, we were able to measure the {\CIV} emission directly with the help of our ICA reconstruction scheme in Rankine+2020. Now, with these measurements, we have connected the emission outflow signatures with the extremely high velocity outflows observed in absorption.

Rankine+2020 has already shown that, except for the top-left quadrant of the {\CIV} emission space, BALQSOs and non-BALQSOs share a large range of the parameter space; in other words, most emission-line properties of BALQSOs are indistinguishable from non-BALQSOs. 
This is not entirely surprising as broad absorption has been found to be variable and it can emerge  \citep[i.e.,][]{Hamann08} and disappear (\citealt{FilizAk12}) in quasar spectra. 
Quasars with EHVOs, on the other hand, populate the bottom-right quadrant of the {\CIV} space, showing the largest {\CIV} blueshifts; their distinct {\CIV} emission line profiles suggest that EHVOs may be a distinct class of quasars or a natural progression from BALQSOs as the speeds increase. In fact, BALQSOs with large speeds ($V_{\rm max} \gsim$~15,000 {\kms}) and EHVOs show similar properties in the {\CIV} emission space, which suggests that the arbitrary cutoff at 0.1$c$ due to the presence of {\SiIV + {\OIV}] emission should be reconsidered: either placed around $\sim$0.05$c$ or properties might change gradually as velocity increases. Many quasars that show EHVOs also include outflows in the $\sim$0.05$c$-0.1$c$ velocity range. As described in RH+2020, absorption on top of the emission line complex is difficult to confirm, but reconstructions such as those in Rankine+2020 are very promising to increase the number of EHVOs known in this region. Also, more studies comparing properties of EHVOs and those of BALQSOs with large speeds will help us learn whether EHVOs are a distinct class or an extension of large-speed BALQSOs. 

Rankine+2020 showed a correlation between values of bolometric luminosity, Eddington ratio, BALQSOs' speeds and {\CIV} blueshifts. Finding that EHVOs occupy the parameter space with largest {\CIV} blueshifts, we expected they would also show correlations with these parameters. In RH+2020, we had already found that EHVOs tend to have slightly larger bolometric luminosities, which we also find here, but we did not observe originally 
largest values of Eddington ratios. After investigating the different prescriptions to calculate black hole mass (\citealt{Shen11}\footnote{$M_{\rm BH}$ values from \citet{Shen11}  were used in the DR9Q EHVOs in RH+2020}; Rankine+2020), we find that measurements that do not take into account the excess, non-virial, blue emission for quasars with large {\CIV} blueshifts can be largely overestimated at large values (\citealt{Coatman17}). 
Using Rankine+2020 we find more moderate values of black hole mass in EHVOs with an overall smaller spread. This results in a correlation between $L_{\rm bol}$ and Eddington ratio for EHVOs, which also show some of the largest values of these two properties (yellow and light green dots in Figure \ref{LbolMBHRedd}). Quasars at large redshift ($z >$~5.8; \citealt{Bischetti22}) also seem to have larger values of Eddington ratios. 
Our results confirm the increase of sub-relativistic accretion disk winds with increasing Eddington ratios predicted by \citet{Giustini19}. While our study remains in the low end of the Super-Eddington regime, \citet{Giustini19} predicts more polar jets which would be interesting to explore in the EHVO cases with large Eddington ratios.

The dearth of BALQSOs in the top-left quadrant (or with the lowest {\CIV} distances) correlates with stronger {\HeII} $\lambda$1640.42 emission lines. {\HeII} is stronger for harder spectral energy distributions (SEDs; \citealt{Casebeer06}), which result in weaker winds because the gas may be too ionized for radiation line driving to be efficient due to the electrons no longer being bound to the nuclei. Indeed, \citet{Richards11} and \citet{Baskin15} already showed that BAL-type quasars have typically weaker {\HeII} emission in their composite spectra. 
However, Rankine+2020 suggested that for a given quasar, the location in {\CIV} emission space is more important for determining the quasar's properties such as the bolometric luminosity, Eddington ratio, and strength of the ionising SED than whether or not the quasar is a BALQSO. Our results further this conclusion as we find that the location in {\CIV} space is more important for determining the bolometric luminosity, Eddington ratio, and black hole mass than the presence of fast EHVOs as their values for the EHVO sample are consistent with both the non-BAL and BALQSOs. The {\HeII} emission is close to indistinguishable from the parent population; however, the suspected limit on {\HeII} EW for the presence of EHVOs suggests that their is an SED hardness above which EHVOs cannot be driven. This is in contrast to BALQSOs which can have {\HeII} EWs higher than this limit although larger EHVO samples are needed to confirm this.
We also find that the $V_{\rm max}$ values for EHVOs show an incipient trend in the {\CIV} emission space towards larger {\CIV} blueshifts, as BALQSOs show, but it needs to be confirmed by enlarging the sample.

\section{Conclusions}

We have examined the continuum and emission line properties, namely {\CIV} and {\HeII}, bolometric luminosities, Eddington ratios, and black hole masses, of a sample of 31 quasars with EHVOs and considered these properties in the context of the quasar population as a whole. A summary of our findings:
\begin{enumerate}
    \item Typically, EHVOs are hosted by quasars found in the lower-right quadrant of {\CIV} space, with large blueshifts and low EWs (Figure \ref{CIVbluehist}), both more extreme than the average values for all non-BALQSOs and BALQSOs, but similar to those BALQSOs with large speeds.
    
    \item The small sample size deters us from making any substantial claims about velocity trends; however, the maximum EHVO velocity increases with increasing {\CIV} blueshift, consistent with the trends found in the BALQSOs. The minimum velocity on the other hand shows no such trend (Figure \ref{vminvmax}).
    
    \item The bolometric luminosities, black hole masses, and Eddington ratios of our EHVO sample are very similar to the corresponding values of other quasars in the parent sample in the same portion of the {\CIV} emission line parameter space (Figure \ref{LbolMBHRedd}). Bolometric luminosities and Eddington ratios appear to be on the slightly larger realm of the parameter space, similarly to what occurs for high-$z$ quasars. 
    
    \item The {\HeII} EWs, an indicator of the strength of the ionising SEDs, of the EHVO sample are close to identical to the parent sample given their location in {\CIV} space; however, there is the suggestion that their is an upper limit of {\HeII} strength and therefore SED hardness for a quasar to host an EHVO (Figure \ref{HeII_CIVdist}). 
\end{enumerate}

With new samples of EHVO quasars (for example, DR16Q, Rodr\'iguez Hidalgo et al. in prep) we will be able to improve the statistics of the trends we have discovered with EHVO velocities, Eddington ratio, luminosity, and black hole mass. Larger samples will also facilitate further study of potential limits of SED hardness via examination of the {\HeII} emission line.

Finding more potential connections between outflows in emission and absorption will become the cornerstone work for future James Webb Space Telescope (JWST) and Atacama Large Millimeter/submillimeter Array (ALMA) proposals to study [CII] and OIII] emission and observe the host galaxies of EHVO quasars.

\begin{acknowledgments}
We thank Paul Hewett, Gordon Richards, and Pat Hall for their insightful comments. 
PRH acknowledges support from the National Science Foundation Astronomy \& Astrophysics (award  2107960). ALR acknowledges support from UKRI (grant code: MR/T020989/1). 

Funding for SDSS-III has been provided by the Alfred P. Sloan Foundation, the Participating Institutions, the National Science Foundation, and the U.S. Department of Energy Office of Science. The SDSS-III website is http://www.sdss3.org/.
SDSS-III is managed by the Astrophysical Research Consortium for the Participating Institutions of the SDSS-III Collaboration including the University of Arizona, the Brazilian Participation Group, Brookhaven National Laboratory, Carnegie Mellon University, University of Florida, the French Participation Group, the German Participation Group, Harvard University, the Instituto de Astrofisica de Canarias, the Michigan State/Notre Dame/JINA Participation Group, Johns Hopkins University, Lawrence Berkeley National Laboratory, Max Planck Institute for Astrophysics, Max Planck Institute for Extraterrestrial Physics, New Mexico State University, New York University, Ohio State University, Pennsylvania State University, University of Portsmouth, Princeton University, the Spanish Participation Group, University of Tokyo, University of Utah, Vanderbilt University, University of Virginia, University of Washington, and Yale University. 
\end{acknowledgments}

\software{astropy \citep{Python2013},  
          {\CIV} distance \citep{CIVdist}
          }

\newpage

\bibliography{bibliography.bib}
\bibliographystyle{apj_modified}

\end{document}